\documentclass[conference]{IEEEtran}
\IEEEoverridecommandlockouts

\def\BibTeX{{\rm B\kern-.05em{\sc i\kern-.025em b}\kern-.08em
    T\kern-.1667em\lower.7ex\hbox{E}\kern-.125emX}}

\usepackage{cite}
\usepackage{amsmath,amssymb,amsfonts}
\usepackage{algorithmic}
\usepackage{textcomp}
\usepackage{xcolor}

\usepackage{booktabs}
\usepackage{verbatim}
\usepackage{graphicx}
\usepackage{caption}
\usepackage{subcaption}

\newtheorem{definition}{Definition}
\newtheorem{proposition}{Proposition}

\newcommand{\btc}{Bitcoin}
\newcommand{\op}{BitcoinF}
\newcommand{\ede}{\emph{$\epsilon$-Expected Dominant Strategy Equilibrium}}
\newcommand{\ene}{\emph{$\epsilon$-Expected Nash Equilibrium}}

\newcommand{\fm}{\emph{greedy}}
\newcommand{\fifo}{FIFO}
\newcommand{\qonen}{FM}
\newcommand{\qtwon}{FIFO}

\begin{document}

\title{\op: Achieving Fairness for Bitcoin in Transaction-Fee-Only Model}

\author{\IEEEauthorblockN{Shoeb Siddiqui, Ganesh Vanahalli, Sujit Gujar}
\IEEEauthorblockA{\textit{International Institute of Information Technology}\\
Hyderabad, India \\
shoeb.siddiqui@research.iiit.ac.in, ganesh.vanahalli@students.iiit.ac.in, sujit.gujar@iiit.ac.in}
}

\maketitle

\begin{abstract} 
A blockchain, such as \btc, is an append-only, secure, transparent, distributed ledger. A fair blockchain is expected to have healthy metrics; high honest mining power, low \emph{processing latency}, i.e., low wait times for transactions and stable \emph{price of consumption}, i.e., the minimum transaction fee required to have a transaction processed. As Bitcoin matures, the influx of transactions increases and the block rewards become insignificant. We show that under these conditions, it becomes hard to maintain the \emph{health of the blockchain}.
In Bitcoin, under these \emph{mature operating conditions} (MOC), the miners would find it challenging to cover their mining costs as there would be no more revenue from merely mining a block. It may cause miners not to continue mining, threatening the blockchain's security. Further, as we show in this paper using simulations, the cost of acting in favor of the health of the blockchain, under MOC, is very high in Bitcoin, causing all miners to process transactions greedily. It leads to \emph{stranded transactions}, i.e., transactions offering low transaction fees, experiencing unreasonably high processing latency. To make matters worse, a compounding effect of these stranded transactions is the rising price of consumption. Such phenomena not only induce unfairness as experienced by the miners and the users but also deteriorate the health of the blockchain.

We propose \op\ transaction processing protocol, a simple, yet highly effective modification to the existing \btc\ protocol to fix these issues of unfairness. \op\ resolves these issues of unfairness while preserving the ability of the users to express urgency and have their transactions prioritized.

\end{abstract}

%%%%%%%%%%%%%%%%%%%%%%%%%%%%%%%%%%%%%%%%%%%%%%%%%%%%%%%%%%%%%%%%%%%%%%%%%%%%%%%%%%%%%%%%%%%%%%%%%%%%%%%%%

%%%%%%%%%%%%%%%%%%%%%%%%%%%%%%%%%%%%%%%%%%%
\section{Introduction}
\label{sec:intro}
%%%%%%%%%%%%%%%%%%%%%%%%%%%%%%%%%%%%%%%%%%%

Blockchain, introduced in Bitcoin \cite{nakamoto2008bitcoin} by Nakamoto, is an append-only, secure, transparent, distributed ledger, storing data in blocks connected through immutable cryptographic links, with each block extending exactly one previous block.
In blockchain technology, the \emph{miners} validate transactions that, the \emph{users} \emph{publish} (create and broadcast). Miners add valid transactions into the next block(s). Different miners attempt to publish (create and broadcast) the next block. In \emph{Proof of Work} (PoW) blockchains, such as \btc, the miner who solves a cryptographic puzzle first is whose published block is accepted as the extension. Each miner has a different puzzle, yet of the same level of difficulty, which needs computations to solve.

In Bitcoin, there are two types of rewards offered to the miners: \emph{block rewards} and \emph{transaction fees}. Block rewards are incentives that the miners are allowed to pay to themselves, minting currency in every block mined, regardless of the contents of the block. Transaction fees, on the other hand, are incentives offered by the users to the miners to prioritize their transactions. In Bitcoin and similar PoW blockchains, the miners invest resources, such as electricity and hardware, in such computations in anticipation of these rewards. 

It is due to the investment on the part of the miners that they benefit from the \emph{health of the blockchain}.
In the context of this paper, we characterize the \emph{health of a blockchain} by (i) the fraction of mining power held by honest nodes; higher, the better (ii) \emph{processing latency}: one of the performance parameters of the blockchain; lower the better, and (iii) the \emph{price of consumption}; lower the variance, the better.
All of these three metrics are linked to the perceived value of the blockchain and its currency. If the health of the blockchain is good, it is prudent to say that the underlying crypto-currency possesses a good value.

One of the key differences between traditional currency and \btc\ is inflation control. To control inflation, block rewards are halved every four years in \btc. Over time, a scenario develops, in which block rewards are negligible, and the only incentive for the miners is the transaction fee, i.e., the \emph{transaction-fee-only model} (TFOM).

When the block rewards are high in value, we say the \btc\ protocol satisfies \emph{individual fairness for the miners} as it is believed that the current block rewards at least cover the marginal costs of mining blocks. Since the miners are not hard-pressed for revenue, they can include transactions for free in the order that they arrive. This not only ensures that no transaction is stranded but also ensures that the price of consumption does not rise.  A blockchain ecosystem is \emph{fair for the users} if it has (i) low processing latency and (ii) stable price of consumption. Currently, the Bitcoin ecosystem is fair to the miners and users. The vital question we study is, do these three notions of fairness carry forward to TFOM? 

The authors in \cite{carlsten2016instability} showed that in TFOM under low \emph{influx} (incoming volume of transactions), the rational miners will \emph{undercut} instead of following default strategy. While this analysis considers the impact of rational miners in TFOM w.r.t. \emph{forking}, it does not consider the processing latency and the price of consumption.

\begin{figure}
\centering
\includegraphics[scale=0.3]{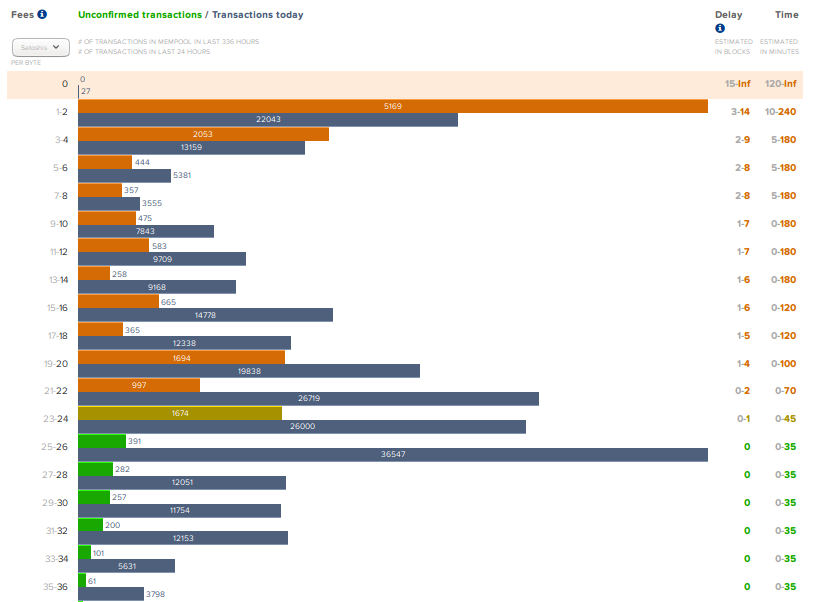}
\caption{Transaction Fees vs Processing Latency \cite{bitcoinfees}} \label{fig:btfees}
\end{figure}

In this paper, our goal is to quantify fairness to the miners and the users and study the impact of TFOM under \emph{standard influx}. Standard influx refers to the case when influx on an average is equal to the maximum outflux (processing capacity) of the blockchain.
The two conditions, TFOM and standard influx, inherently go hand-in-hand as the Bitcoin matures \cite{blockchain.com}. Thus, making it very important to study and contemplate such scenarios. In this paper, we analyze Bitcoin in TFOM and under standard influx, which we term as \emph{mature operating conditions} (MOC), and show that it is unfair for both miners and users. This unfairness, in a nutshell, is exemplified in Fig. \ref{fig:btfees}. We observe that those paying lesser transaction fees are expected to wait upto 9 blocks, and those who pay an insignificant amount of fees are expected to experience a processing latency of 14 blocks.

In TFOM, even when transaction volumes are sufficiently high enough to fill the processing capacity of the blockchain, it is not assured that the miners earn sufficient revenue. 
Insufficient transaction fees can be a major issue, as this puts the blockchain at a security risk due to the possibility of reduced honest mining power, which in turn will deteriorate the health of the blockchain, reducing its value, and hence further dropping of the honest miners.

Miners following \emph{First-In-First-Out} (\fifo) processing ensures low and reasonable processing latency, avoiding stranding transactions entirely. Since no transactions are stranded, the price of consumption does not increase. Hence, miners following \fifo\ help sustain the blockchain's health through good performance and maintaining a stable price of consumption. 
In Bitcoin, under MOC, we show that miners following \fifo\ processing take heavy losses as compared to the ones mining greedily. This is an issue, as miners depend entirely on transaction fees to sustain mining and cannot take considerable losses in order to follow \fifo\ processing. This results in all miners processing transactions greedily. As we show in our analysis, this causes transactions to experience unreasonably high processing latency; such transactions are referred to as \emph{stranded transactions}. This leaves the users with uncertainty about whether or not their transactions will be processed. A compounding effect of stranded transactions is the rising price of consumption.
These issues culminate in unfairness for both the miners and the users. Thus, we say that Bitcoin, in its current form, is unfair under MOC (Proposition \ref{prop:btc}).

We solve these issues of unfairness by proposing a novel protocol, \op, for processing transactions. \op\ enforces a minimum transaction fee and uses two queues, instead of one to process transactions. 
%\op\ ensures fairness under MOC (Proposition \ref{}), while 
\op\ allows the users to express urgency and have their transactions prioritized, just as they can do in Bitcoin. Our game-theoretic analysis proves that the proposed modification to Bitcoin, \op, ensures good health of the blockchain. Thus, we believe \op\ will lead to a stable ecosystem and hence will be fair to the miners and the users (Proposition \ref{prop:op}). While there have been many published works analyzing TFOM, using collected data or using game-theoretic models, to the best of our knowledge, this is the first formal attempt at solving the pressing issues that are bound to arise in TFOM.

\section{Related work}
\label{sec:relim}
There are many papers in the literature studying transaction fees offered in \btc. The authors in \cite{moser2015trends} study the behavior of transaction fees over a period of time, whereas Li et al. \cite{li2018transaction} conduct a theoretical analysis of a queuing game to study the transaction fees, and both conclude that the users paying lesser fee faced higher processing latency. The authors in \cite{houy2014economics}, point out that if transaction fees were to be determined by the free market alone without a block size limit, it would be detrimental to \btc\ as the fees would eventually become zero and miners will no longer have an incentive to mine. Easley et al. \cite{easley2019mining} develop a game-theoretic model, based on observational data, to study transaction fees and explain the behavior of miners and users in equilibrium. The authors in \cite{kroll2013economics} conclude briefly that transaction fees would not play any major role unless the underlying rules of \btc\ are changed. However, \cite{huberman2019economic} suggests otherwise, clearly stating the importance of transaction fees and suggest increasing block rate to check congestion and increase miners' revenue. The authors in \cite{kasahara2016effect} study the effects of transactions paying a small fee and block size limit on the transaction confirmation time using queuing theory. The authors in \cite{koops2018predicting} study how the users tend to offer high fees for their transactions to get included in the block when the demand exceeds block capacity, they model the confirmation time as a particular stochastic fluid queuing process. 

%%%%%%%%%%%%%%%%%%%%%%%%%%%%%%%%%%%%%%%%%%%
\section{Preliminaries}
\label{sec:prelim}
%%%%%%%%%%%%%%%%%%%%%%%%%%%%%%%%%%%%%%%%%%%

In this paper, we focus on Bitcoin when operating under \emph{mature operating conditions} (MOC), i.e., when blocks rewards are negligible (Transaction-fee-only model (TFOM)), and there is standard demand (influx on an average is equal maximum outflux). First, we define all the important terms. Then, we present our model and describe our assumptions. Next, we explain how we simulate miners' behavior and users' behavior.

\subsection{Important Definition}
\label{ssec:imp_defns}
\begin{definition}[Honest miner]
We say a miner is \emph{honest} or \emph{non-adversarial} if it does not willingly attempting to disrupt the Bitcoin ecosystem by adding \emph{invalid transactions} to blocks, attempting to \emph{double spend} or by extending other than the \emph{longest chain}. 
\end{definition}

\begin{definition}[Rational Miner]
We say a miner is \emph{rational} if it; continues mining when \emph{individual fairness for the miners} is guaranteed, acts in favor of the health of the blockchain if the cost of doing so is marginal.
\end{definition}

The act of \emph{processing transactions} simply involves selecting the transactions from the set of received but yet unprocessed transactions, adding them to the block and then publishing the block. This is also known as mining, and it is performed by miners. The system requires a \emph{honest majority} of miners to maintain the security of the blockchain, vis-a-vis \emph{persistence}, against adversarial miners. Persistence is a property that must be ensured by blockchains; It ensures that the \emph{confirmation} (different from processing transactions) of a transaction by an honest node is never disputed by any other honest node. In this paper, we consider that all miners are honest but \emph{rational}. Miners are incentivized to participate in honest mining by rewards. If the miners are not compensated appropriately for their mining efforts, the rational, though honest miners may choose not to mine. Thus reducing the honest power in the network, weakening the blockchain against adversarial attacks, adversely affecting the health of the blockchain. When a miner chooses to stop mining, they are essentially giving up on the value of the blockchain and hence giving up on the significantly high investment they have in it, either in the form of the mining equipment or in the form of the blockchain currency token they hold.

Besides persistence, another property that must be ensured by blockchains is \emph{liveness}. Liveness ensures that a transaction will eventually be processed; however, it is not sufficient as it does not guarantee that transactions will not get \emph{stranded} for a long time, let alone be processed in a reasonable amount of time. As the blockchain technology matures, to maintain competitive performance, a blockchain must ensure that transactions are processed within a reasonable amount of time, i.e., low \emph{processing latency}. Furthermore, another reason to avoid \emph{stranded transactions} is that it leads to increasing \emph{price of consumption}, further deteriorating the \emph{health of the blockchain}.
While stranded transactions can be avoided by simply restricting the amount of transaction fee offered to a single value, this trivial solution is unacceptable as the users' ability to offer a range of fees is required to express urgency and importance in a setting where there is varying demand.

\begin{definition}[Processing Latency]
\emph{Processing latency} refers to the duration, which we measure in terms of blocks, between a user publishing a transaction and a miner processing it (publishing a block containing it).
\end{definition}

\begin{definition}[Stranded Transactions]
\emph{Stranded transactions} are those transactions, that experience unreasonably high processing latency ( $>100$ blocks).
\end{definition}

\begin{definition}[Price of Consumption]
The price of consumption is the minimum transaction fee, as perceived by the users, that must be paid for the transaction to be processed.
\end{definition}

\begin{definition}[Health of a Blockchain]
\emph{Health of a blockchain} is characterized by (i) the fraction of mining power held by honest nodes (ii) \emph{processing latency} and (iii) the \emph{price of consumption}.
\end{definition}

For the security of the blockchain, mainly Bitcoin, more than 50\% of miners should be honest; higher, the better. This is an essential aspect for the system to be fair to both the honest miners and the users, as the decentralized nature of the system depends on it. Processing latency, one of the performance parameters of the blockchain, is the time taken to process a transaction. It is crucial to the blockchain's usability and adopt-ability. It also ensures that transactions are not \emph{stranded} in the system and are added to the blockchain within a reasonable time. The price of consumption is the minimum transaction fee that must be paid for the transaction to be processed; for stability; lower the variance, the better.

The health of the blockchain would be better if more miners are honest. Given the miner's stake in the ecosystem they participate in, it is fair to say that they would act in favor of the health of the blockchain as long as the cost of doing so is marginal.

\begin{definition}[Individual Fairness for The Miners]
We say the given blockchain protocol satisfies \emph{individual fairness for the miners} if the rewards from a block are at-least the cost of mining it.
\end{definition}
\begin{definition}[Fairness of The Users]
We say the given blockchain protocol satisfies \emph{fairness for the users} if the users experience\\
(i) reasonable processing latency,  \\
(ii) stable price of consumption (i.e., $f_{min}$), and \\
(iii) decreasing average processing latency with increasing $\eta$.
\end{definition}

\begin{definition}[Fair Blockchain]
We say that the blockchain ecosystem is fair if it satisfies individual fairness for the miners and fairness for the users. 
\end{definition}

Since acting in the blockchain's favor yields optimal results required for the blockchain to perform competitively, it is imperative that miners do not deviate from it. Deviations from \fifo\ processing can not be clearly detected and hence can not be actively discouraged; the only solution is to create an environment that ensures that the miner cannot benefit from such deviations. We quantify this as a game-theoretic equilibrium.
Let $\mathcal{M}=\{m_1,m_2,\ldots,m_k\}$ be the set of miners and $S$ be the set of strategies for the miners in the blockchain to act upon. 
\begin{definition}[\ede]
We say $s=(s_{m_1}^*,s_{m_2}^*,\ldots,s_{m_k}^*), s_{m_i}^* \in S$ is \emph{\ede} for the miners if for all the miners, 
\begin{footnotesize}
\begin{align*}\mathbb{E} f_{txn}(s'_{m_i},s_{-m_i}) < (1+ \epsilon) \mathbb{E} f_{txn}&(s_{m_i}^*,s_{-m_i}) \\ & \forall s_{-m_i} \in S_{-m_i},\, \forall{m_i \in \mathcal{M}}
\end{align*}
\end{footnotesize}
where $s'_{m_i} \neq s_{m_i}^*$. $s_{-m_i}$ and $S_{-m_i}$, indicate the strategy profile and set of strategy profiles, followed by the miners except $m_i$. The expectation is w.r.t. randomness in influx of the transactions and the variance in the transaction fees.
\end{definition} 
The above definition may be too strong and difficult to achieve. Hence, we also work with the following, a weaker equilibrium concept from game theory.

\begin{definition}[\ene]
We say $s^*=(s_{m_1}^*,s_{m_2}^*,\ldots,s_{m_k}^*)$, $s_{m_i}^*\in S$, is \emph{\ene} for the miners, if for each miner, the expected revenue per block by following any strategy is not more than $(1+\epsilon)$ times what it would have obtained by following $s_{m_i}^*$; provided the other miners are following $s_{-m_i}^*$. I.e., 
\begin{footnotesize}
\begin{equation*}
    \mathbb{E} f_{txn}(s'_{m_i},s_{-m_i}^*) \leq (1+\epsilon)\mathbb{E} f_{txn}((s_{m_i}^*,s_{-m_i}^*))\;\forall s'_{m_i} \in S, \;\forall{m_i \in \mathcal{M}}
\end{equation*}
\end{footnotesize}
where, $s_{-m_i}^*$ indicates the strategy profile in $s^*$ by the miners except $m_i$. The expectation is w.r.t. randomness in influx of the transactions and the variance in the transaction fees. 
\end{definition}

%%%%%%%%%%%%%%%%%%%%%%%%%%%%
\subsection{Model}
\label{ssec:ana_model}
%%%%%%%%%%%%%%%%%%%%%%%%%%%%

For a tractable analysis, we simulate the execution of the Bitcoin protocol in steps of 10 mins, i.e., each block is published (created and broadcast) every 10 mins. Typically, the time duration between two consecutive blocks is random, with the expected value of 10 mins. We assume that each block can contain a maximum $bs_{max}$ number of transactions. It is important to note that the maximum size of a block in \btc\ is an inherent limitation of the Bitcoin protocol \cite{rizun2015transaction,bitcoin_magazine_2019}. In our analysis, we assume, in accordance with the standard influx condition under MOC, the number of transactions that arrive in each step is \emph{not} constant and follow a Poisson distribution with mean $bs_{max}$. A Poisson distribution is prudent here as the influx is a discrete number of events (w.r.t. arriving transactions) occurring in step, which is a fixed interval of time, with a known constant rate of $bs_{max}$.  Hence, the average influx (in terms of the number of transactions) is equal to the maximum outflux (i.e., maximum block size). Here, \emph{outflux} is the transactions that are processed by the miners.

We split the transaction fee offered by the user as, $f_{txn} = f_{min} + f_{extra}$. $f_{min}$ is the minimum amount of fee, as perceived by the user, that must be included for the miner to process the transaction; it reflects the price of consumption. $f^0_{min} \leq f_{min}$, is the minimum transaction fee set by the protocol, and hence it is the initial value of $f_{min}$ that the users start with. As there is no restriction on the minimum $f_{txn}$ in Bitcoin, $f^0_{min}$ is $0$ as per the protocol. The $f_{extra}$ is the extra fee the user would like to give to the miners to prioritize their transaction over others.

We model the aggression of a user towards $f_{extra}$ through a parameter  $0\leq\eta<\infty$; higher the $\eta$, higher the $f_{extra}$. Each user, when publishing the transaction, calculates the amount of extra fee they would like to pay as a monotonically increasing function, $f_{extra}=\phi(\eta)$. Let $\psi_{\lambda}(\eta)$, a \emph{pdf} characterized by a static parameter $\lambda$, be the distribution that captures the fraction of users having aggression level towards $f_{extra}$ as $\eta$.

During each step, transactions are collected by the miners. At the end of each step, the miners form a block out of the currently unprocessed transactions followed by ``instantly'' publishing it. This merely a simplification of the process of continually updating the block while attempting to solve the cryptographic puzzle along with the assumption that a block would be mined by the end of the step.

We consider two modes that the miners may operate in; i) \fm\ processing, where miners greedily include transactions, i.e., they include the highest \emph{or} lowest valued transactions, whichever may be more profitable ii) \fifo\ processing, where miners process transactions on a \emph{First-In-First-Out} basis. These two modes of miner operation form the strategy space of the miners. Since the influx, as well as outflux, is in terms of the number of transactions, it is understood that all transactions are considered to be of the same size in terms of the space taken on the block. Hence \fm\ processing only considers the transaction fee offered and not the size of the transaction in bytes.

\subsubsection*{Assumptions on Miners' Behaviour}
We assume that all miners are honest but rational. The rationality of the miners implies the following:
\begin{itemize}
    \item Miners would continue to mine and sustain the blockchain, as long as mining costs are covered.
    \item Since the health of the blockchain is crucial to the value miners obtain from mining, and that all miners inherently understand this, miners act in favor of sustaining the health of the blockchain, i.e., follow \fifo\ processing, if the cost of doing so is marginal.
\end{itemize}

\subsubsection*{Assumptions on Users' Behaviour}
We assume that the user would like to have his transaction processed within a reasonable time and that their inclination to have their transaction prioritized is characterized by their aggression level, $\eta$ towards paying a higher $f_{extra}$. This assumption implies the following: 
\begin{itemize}
    \item The user chooses $f_{txn}$ by first considering the $f_{min}$ and then deciding $f_{extra}=\phi(\eta)$ where $\eta$ captures its aggression parameter. Note that, the system need not know the $\eta$ for each individual user, but for analysis, we use the distribution of users against $\eta$ (i.e., $\psi_{\lambda}(\eta)$ is known).
    \item The user observes transactions, below a certain threshold of fees, being stranded, they concede to making the presumption that this threshold of fees is the new $f_{min}$, in the sense that transaction below this fees will not be processed in a reasonable time. This is because a user will not attempt to publish a transaction if they do not expect it to be processed. 
\end{itemize}

\subsection{Simulation Setup}
\label{ssec:simsetup}
Since the theoretical analysis is intractable in such a complex scenario. We use simulations to support our arguments. For a fair and consistent analysis, we use the same simulation setup and parameters to simulate both Bitcoin as well as our protocol. All simulation results were averaged over $10$ runs.

As mentioned in Section \ref{ssec:ana_model}, the execution of the protocol in consideration, proceeds in steps. Each step represents the time between blocks. At the end of each step, a block is added to the chain. Each block has a maximum capacity, $bs_{max}=1000$, in terms of transactions. All the simulations are run in the setting where the average influx is Poisson distributed with mean equal to $bs_{max}$.

We take, $\psi_{\lambda}(\eta) = \lambda \cdot e^{- \lambda \cdot \eta}$, an \emph{exponential distribution} characterized by $\lambda=3$, where $\lambda$ is the rate parameter of the exponential distribution.
$\phi$, the function used to calculate the $f_{extra}$ a user with aggression parameter $\eta$ gives, we take to be $\phi(\eta)=f_{extra}=e^{\eta}-1$.

We take the granularity of the size of miners in terms of their mining power to be $\delta=0.05$; and $0\leq\beta\leq1$ to be the fraction of miners that follow \fm, while the rest follow \fifo.

To observe the cost of mining as per \fifo\ as opposed to acting \fm, and further investigate the establishment of equilibrium, we simulate both, \op\ and \btc\ with varying values of $\beta$. We then consider the resulting \emph{average revenue per block} mined for both \fifo\ and \fm\ behavior, in both \op\ and \btc. The fraction of mining power controlled is represented in the simulation by setting the same fraction as the probability of mining a block.

To emulate the user's characteristic of observing the change in $f_{min}$ based on the stranded transactions, we use epochs of observation. Each \emph{observational epoch} is a series of subsequent steps at the end of which the users change their $f_{min}$ accordingly. At the end of each epoch, the users check the average processing latency of transactions that were published in this epoch. The highest transaction fee that, experiences an average processing latency high enough to be considered stranded, is considered to be the new $f_{min}$. The length of the observational epoch, i.e., the number of steps the users observe after which they change their presumption of $f_{min}$, we take to be $1000$ steps. We use $100$ to be limit to the processing latency in units of steps, after which we consider a transaction to be stranded.

Now we study \btc\ protocol in the next section. 
%%%%%%%%%%%%%%%%%%%%%%%%%%%%%%%%%%%%%%%%%%
\section{Bitcoin Transaction Processing: Analysis under MOC}
\label{sec:}
%%%%%%%%%%%%%%%%%%%%%%%%%%%%%%%%%%%%%%%%%%
First, we describe \btc\ protocol, and then explain what assumptions we make, highlight the specifics of \btc\ simulation, and the inference from the simulations.
\subsection{Bitcoin Protocol}
In Bitcoin, the market for ``space on the block'' is a completely free market (FM), there is no regulation on how the transactions must be processed or how much transaction fee must be given. 
In Bitcoin, a block contains only one section (we refer to this as the \qonen\ section) where there are no restrictions (except our assumption on the maximum number of transactions per block, i.e., block size). The users publish only one instance per transaction containing $f_{txn}=f_{min}+f_{extra}$. Initially, $f_{min}=0$ as the Bitcoin protocol does not state any minimum transaction fee that must be included.

In TFOM, each miner must collect transaction fees to sustain their mining efforts. Even when assuming an influx of transactions that is on an average sufficient to fill the blocks, there is no guarantee that the incoming transactions will contain sufficient transaction fees to sustain the mining efforts. Hence, individual fairness for the miners can not be established.

In Bitcoin, while the block rewards alone are sufficient to sustain mining efforts, the rational miners can be expected to process transactions in a \fifo\ manner, especially since the number of users offering a competitive fee is minimal. However, under MOC, we assume in our analysis that when processing transactions from the \qonen\ queue to be added to the \qonen\ section of the block, the miners pick the transactions offering highest transaction fees, as shown in Fig. \ref{fig:btcrep}. The miners could choose to follow \fifo\ while processing transactions from the \qonen\ queue. They might want to do this to preserve the health of the blockchain. However, we expect our simulation-based game-theoretic investigation into the establishment of equilibrium will yield that the miners lose a large part of their revenue by following \fifo\ processing in the \qonen\ queue. Hence the miners cannot be expected to act in favor of the blockchain's health as the loss is considerable. Thus, we assume the miners will gather transactions greedily.

The users suffer as a consequence of the miners not being able to follow \fifo\ processing. Not wanting to follow \fifo\ would not be cause for concern to the users if the influx of transactions to be processed was low. If the influx of transactions is low, then no matter the transaction fees, all transactions would be included in the next block. In low influx scenarios, the competition is not high, and the users realize that there is no need to pay any more than a marginal fee, as their transaction would get included in the next block regardless. However, as we show in our analysis, issues arise in higher influx scenarios where the users \emph{must} pay competitive fees to have their transactions processed or risk having them stranded. Further, since these stranded transactions are public knowledge, this, as we see in our analysis, causes the price of consumption to rise, adversely affecting the health of the blockchain.

In Bitcoin, under MOC, as we show using simulations, there are a large number of transactions that get stranded. This causes the users to be uncertain about when their transaction will be processed, or even if it ever will be. These stranded transactions are public knowledge. Thus, over a sufficiently long series of consecutive steps, if it is observed that there are transactions that are stranded, the users are likely to treat the highest transaction fees offered by the stranded transactions as the new $f_{min}$ to avoid their transactions being pushed further down in the queue. Given this new $f_{min}$, the phenomenon will repeat. Depending on the observations of stranded transactions, $f_{min}$ can increase, decrease (not below $f^0_{min}$), or remain constant. We study what is likely happen for $f_{min}$ \btc\ via simulations.

\subsection{Simulation Specifics}
Since the miners only include transactions greedily, to emulate miners in this scenario, we have only one section (\qonen\ section) of the block, which spans the entire capacity of the block, to which we keep adding the highest valued transaction till the block is full. As per the Bitcoin protocol, we keep the initial $f_{min}=0$.

First, we conduct simulations to investigate the establishment of equilibrium, which we expect to state that to follow \fm\ processing is \ede\ validating the assumption that the miners follow \fm\ processing from the \qonen\ queue. We assume the same in our simulation estimating average processing latency and the change in $f_{min}$ with observation epochs.

\subsection{Simulation Results and Inference}

First, we discuss the result of our investigation of equilibrium of miner behavior in the \btc\ ecosystem.
The strategy space of the miners here is $S = \{\fifo,\fm\}$, where $s\in S$ is the mode of processing (as in Section \ref{ssec:ana_model}) of the miner in the \qonen\ queue.

\begin{figure*}[t]
\begin{subfigure}{0.50\columnwidth}
\centering
\includegraphics[width=1\columnwidth]{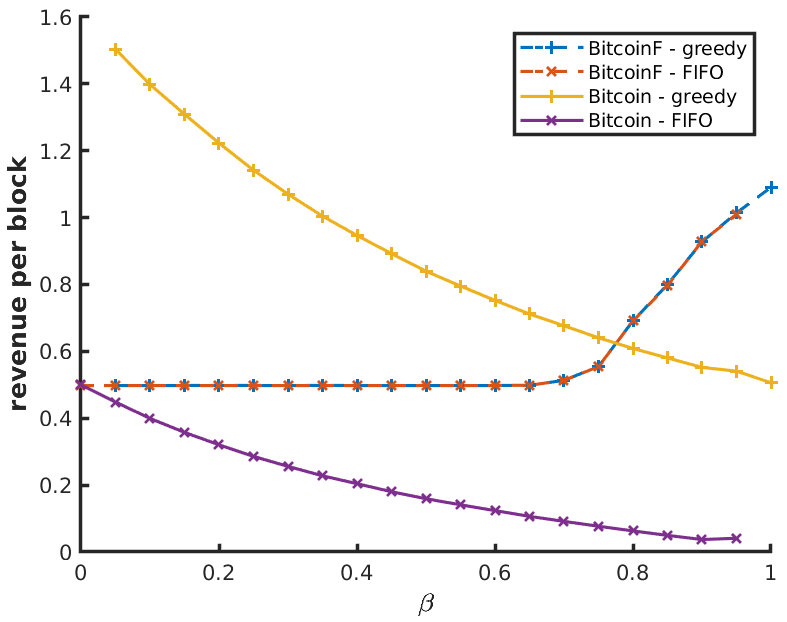}
\caption{Difference in revenue of \fifo\ and \fm} \label{fig:finalrevdiffvsbeta}
\end{subfigure}
%\hfill
\begin{subfigure}{.50\columnwidth}
\centering
\includegraphics[width=1\columnwidth]{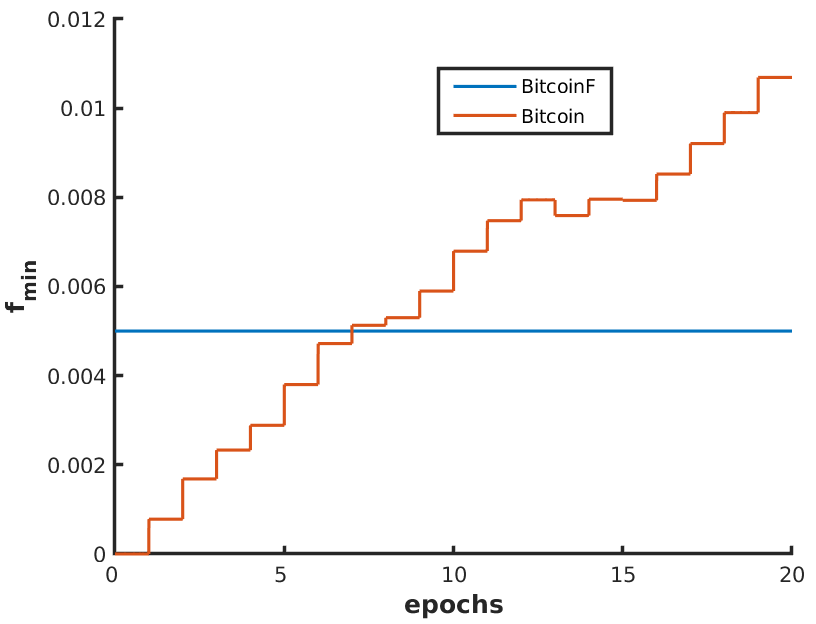}
\caption{$f_{min}$ vs steps} \label{fig:btcbu}
\end{subfigure}
%\hfill
\begin{subfigure}{.50\columnwidth}
\centering
\includegraphics[width=1\columnwidth]{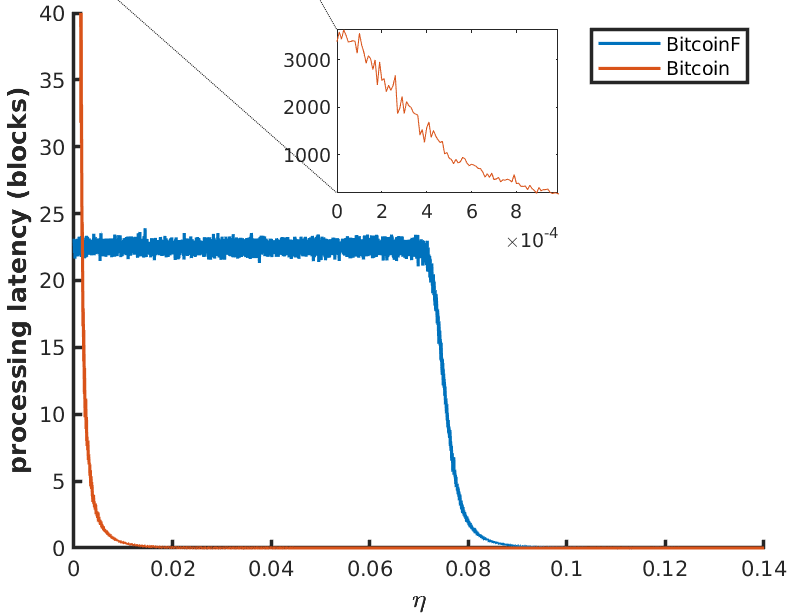}
\caption{Processing latency vs $\eta$} \label{fig:plvseta}
\end{subfigure}
%hfill
\begin{subfigure}{0.50\columnwidth}
\centering
\includegraphics[width=1\columnwidth]{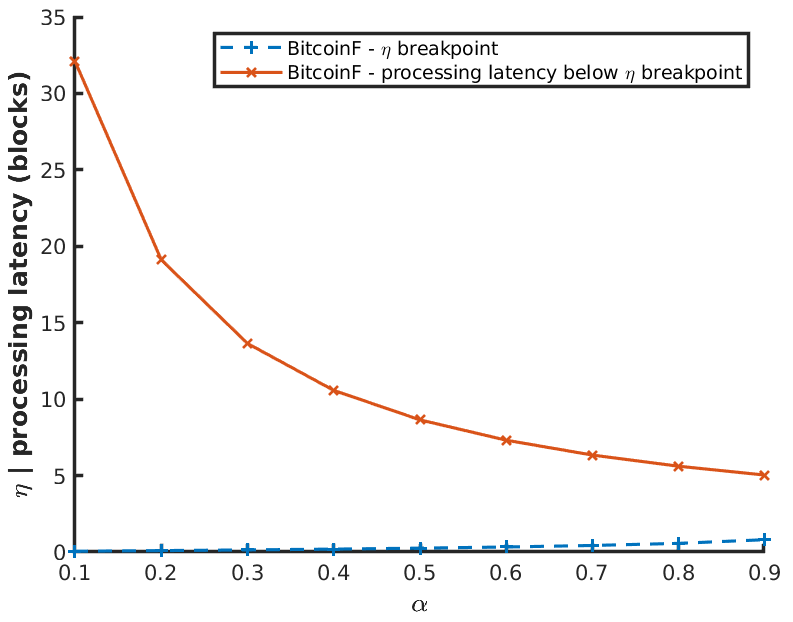}
\caption{Trend w.r.t. alpha} \label{fig:alphatrend}
\end{subfigure}
\caption{Simulation Results}
\end{figure*}

Clearly, from Fig. \ref{fig:finalrevdiffvsbeta}, we see that:
\begin{footnotesize}
\begin{align*}\mathbb{E} f_{txn}(s'_{m_i},s_{-m_i}) < (1+ \epsilon) \mathbb{E} f_{txn}&(s_{m_i}^*,s_{-m_i}) \\ & \forall s_{-m_i} \in S_{-m_i},\, \forall{m_i \in \mathcal{M}}
\end{align*}
\end{footnotesize}
where $s'_{m_i} = \fifo$, $s_{m_i}^* = \fm$ and $\epsilon=0$.

Intuitively, miners make significantly higher revenue if they followed \fm\ processing as opposed to \fifo\ processing regardless of $\beta$. Thus we say that following \fm\ processing in the \qonen\ queue is \ede\ with $\epsilon=0$. 

Confirming our assumption about miner behavior, we now discuss the results of our simulations regarding the fairness of the blockchain. As we can see from Fig. \ref{fig:btcbu}, the $f_{min}$ rises with observational epochs. This causes instability in the price of consumption. The figure also implies that when the influx has not been \emph{standard} for long enough, Bitcoin cannot ensure that mining costs will be covered, i.e., individual fairness for miners is not guaranteed. In Bitcoin, Fig. \ref{fig:plvseta} shows that, the users that have very little aggression towards paying $f_{extra}$, experience unreasonably high processing latency, i.e., stranded transactions. These stranded transactions, in turn, cause the price of consumption to rise, as seen in Fig. \ref{fig:btcbu}.

These observations are summarized as Proposition \ref{prop:btc}.

\begin{proposition}
\label{prop:btc}
If the miners strategy space is $S=\{\fifo, \fm\}$, it is \ede\ for the miners to follow \fm\ with $\epsilon=0$. As a consequence, the \btc\ ecosystem is not a fair under MOC.
\end{proposition}

\begin{figure}[t]
\begin{subfigure}{0.49\columnwidth}
\centering
\includegraphics[width=1\columnwidth]{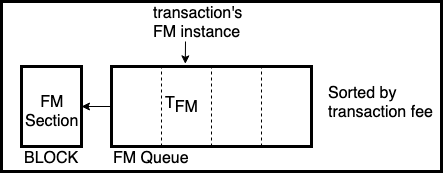}
\caption{\btc}
\label{fig:btcrep}
%hfill
\end{subfigure}
\begin{subfigure}{0.49\columnwidth}
\centering
\includegraphics[width=1\columnwidth]{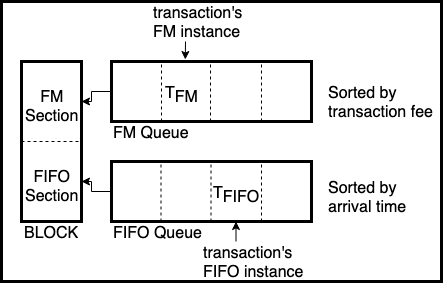}
\caption{\op}
\label{fig:oprep}
\end{subfigure}
\caption{Transaction processing in \btc\ and \op}
\end{figure}

\begin{comment}
\begin{figure}%{0.67\columnwidth}
\centering
\includegraphics[width=0.7\columnwidth]{alphatrend.png}
\caption{Trend w.r.t. alpha} \label{fig:alphatrend}
\end{figure}
\end{comment}

\section{Achieving Fairness under MOC}
As discussed in the previous section, under MOC, \btc\ faces challenges. To resolve this, we propose a simple modification to \btc, which we call \op. 
\subsection{\op}
To solve the issues of unfairness for both the miners and the users, we propose a protocol to process transactions.
Our approach is two-fold: Firstly, we enforce a minimum fee of $f^0_{min}(>0)$ that is to be included in every transaction. Secondly, we introduce a section in the block that only accepts transaction instances with $f_{txn}=f^0_{min}$, called the \qtwon\ section of the block. So now, there are two sections in the block: \qonen\ and \qtwon. The \qtwon\ section has a size of $\alpha \cdot bs_{max}$, whereas \qonen\ has the remaining. This is illustrated in Fig. \ref{fig:oprep}. In our simulation of \op, we set $\alpha=0.2$.

Formally, we propose \op\ as a \emph{block validation rule}. This rule will have two parameters, $\alpha$, and $f^0_{min}$. Miners shall only accept and extend blocks that follow the rule. We expect that when this protocol is implemented, it will be enforced by the honest miners, as is commonplace in blockchain ecosystems.

\begin{definition}[\op: Block Validation Rule]
Each block must contain $\alpha \cdot bs_{max}$ transactions offering $f_{txn}=f^0_{min}$.
\end{definition}

A typical execution is described as follows:
\begin{itemize}
    \item Users when they want to add a transaction to the blockchain broadcast two instances of the same transaction; one instance that has $f_{extra}=0$, and the other instance where $f_{extra}$ is as chosen by the user. Both instances must include at least $f^0_{min}$ as required.
    \item The miners collect these instances of every transaction and add the instance with $f_{extra}=0$ to the \qtwon\ queue and the other instance, the one with $f_{extra}$ to the \qonen\ queue.
    \item When an instance of a transaction is processed, the other instance is invalidated.
    \item The block can have transactions in the \qonen\ section \emph{only} if the \qtwon\ section of the block is completely filled. The honest but rational miners naturally would first add as many transactions as possible to the \qonen\ section from the \qonen\ queue (they may do this however they please, but naturally they choose the ones with highest transaction fees), while keeping aside sufficient transactions to fill the \qtwon\ section of the block. Then, the miners fill the \qtwon\ section of the block with transactions selected from the \qtwon\ queue in a \fifo\ manner.
\end{itemize}

When processing transactions from the \qonen\ queue to be added to the \qonen\ section of the block, the miners naturally pick the transactions offering the highest fees. Further, we assume that while processing transactions from the \qtwon\ queue, the miners follow \fifo.

The miners could choose to follow \fm\ processing, i.e., add the minimum valued transactions instead of following \fifo\ processing to fill the \fifo\ section. They might want to do this, as \fifo\ processing might process some slightly higher valued transactions through the \fifo\ section of the block, voiding the $f_{extra}$ it offers; thus, by following \fm\ processing, the miners process the least valued transactions through the \fifo\ section, while keeping the slightly higher valued transactions (which would otherwise get processed through \fifo) for later, to be processed through the \qonen\ section of the block. However, we expect our simulation-based game-theoretic investigation into the establishment of equilibrium will yield that the miners gain a negligible profit by following \fm\ processing as opposed to \fifo\ processing in the \qtwon\ queue. Hence the miners can be expected to act in favor of the blockchain's health as the loss is insignificant. Thus, we assume the miners will follow \fifo\ processing in the \qtwon\ queue.

\subsection{Simulation Specifics}
The size of the \qtwon\ section is set to be $\alpha \cdot bs_{max} = 200$ transactions, to which transactions are added in a \fifo\ manner. The size of the \qonen\ section is set to be $(1-\alpha) \cdot bs_{max}=800$ transactions, to which transactions are added greedily. The \qonen\ queue is processed before the \qtwon\ queue, as is the expected behavior of the miners. To emulate the random order of the transactions' arrival during a step, when processing from the \qtwon\ queue, random transactions are picked from the set of transactions with the highest processing latency. As per the protocol, initially, the value of $f^0_{min}$ is set appropriately to compensate miners, we, in our simulation, set it to be $0.005$.

First, we conduct a simulation to investigate \ene\, which we expect to state that to follow the \fifo\ is \ene\ validating the assumption that the miners follow \fifo\ processing from the \qtwon\ queue. We assume the same in our simulation estimating average processing latency and the change in $f_{min}$ with observation epochs.

\subsection{Simulation Results and Inference}

First, we discuss the result of our investigation of equilibrium of miner behavior in the \op\ ecosystem.
The strategy space of the miners here is $S = \{\fifo,\fm\}$, where $s\in S$ is the mode of processing (as in Section \ref{ssec:ana_model}) of the miner in the \qtwon\ queue.

Clearly, from Fig. \ref{fig:finalrevdiffvsbeta}, we see that: \begin{footnotesize}
\begin{equation*}
    \mathbb{E} f_{txn}(s'_{m_i},s_{-m_i}^*) \leq (1+\epsilon)\mathbb{E} f_{txn}((s_{m_i}^*,s_{-m_i}^*))\;\forall s'_{m_i} \in S, \;\forall{m_i \in \mathcal{M}}
\end{equation*}
\end{footnotesize}
where $s^*=\{\fifo,\fifo,\ldots,\fifo\}$ and $\epsilon=0.00037$.

Intuitively, miners gain negligible profit if they followed \fm\ processing as opposed to \fifo\ processing in the \qtwon\ queue when $\beta=0$. Thus we say that all miners following \fifo\ processing in the \qtwon\ queue is \ene\ with $\epsilon=0.00037$.

Here, \ene\ is clearly a much weaker property than the one established by \op. Motivated by the analysis in \cite{zou2010tolerable}, where the authors consider a fraction of agents (our case miners) following honest strategy and remaining agents follow greedy strategy, it is easy to see that our protocol exhibits \ede\ with $\epsilon=0.00037$ when $\beta \leq 0.55$. The protocol does not establish \ede\ for all $\beta$, as after a point, the fraction of miners following \fifo\ processing in the \fifo\ queue drops low enough that transactions start to get stranded, raising the price of consumption and hence raising the average revenue of all miners.

Confirming our assumption about miner behavior, we can now discuss the results of our simulations regarding the fairness of the blockchain. As seen in Fig. \ref{fig:btcbu}, $f_{min}$ remains constant with time. This implies a stable price of consumption. Since the $f^0_{min}$ can be set as per requirements to cover mining costs, \op\ guarantees individual fairness for miners under MOC.
Fig. \ref{fig:plvseta} shows that in \op, no matter the users' aggression towards paying $f_{extra}$, the users experience reasonable processing latency. Since there are no stranded transactions, the price of consumption does not rise, as seen in Fig. \ref{fig:btcbu}. In fact, under MOC, the users would know what processing latency to expect as soon as they publish the transaction. If their $\eta$ is higher than a certain \emph{$\eta$ break-point}, then their transaction will get processed almost immediately; if not, they know the upper bound on the processing latency they will experience. This trade-off between the $\eta$ break-point and processing latency below the $\eta$ break-point as a function of $\alpha$ is visualized in Fig. \ref{fig:alphatrend}. 
These simulation based observations can be summarized as Proposition \ref{prop:op}.
\begin{proposition}
\label{prop:op}
If the miners strategy space is $S=\{\fifo, \fm\}$, it is \ene\ for the miners to follow \fifo\ with $\epsilon=37*10^{-5}$. As a consequence, \op\ ecosystem is a fair under MOC.
\end{proposition}

\subsection{Security Analysis}
\label{sec:opattack}
A strategic and intelligent miner could attempt to use our protocol to leverage an unfair advantage. In this section, we show that such attempts are not quite effective and hence, do not threaten our protocol.

\paragraph{Ignoring the \qtwon\ Section of the block} The miners would ideally like to ignore all the $f^0_{min}$ instances of the transactions. The miners cannot do this as any valid block requires that the \qtwon\ section of the block must be filled entirely before any transactions are added to the \qonen\ section of the block. Also, all transaction instances in the \qtwon\ section are only supposed to offer $f^0_{min}$ fee.

\begin{figure}
\centering
\includegraphics[scale=0.3]{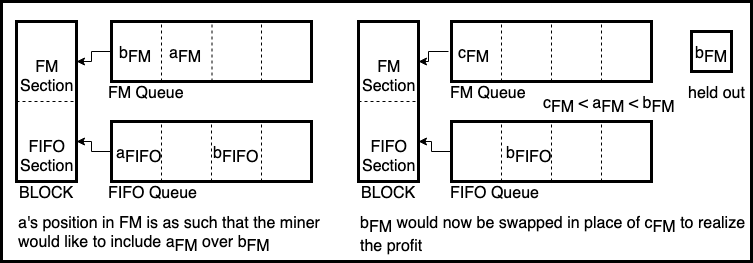}
\caption{Depiction of swapping} \label{fig:swap}
\end{figure}

\paragraph{Swapping in transactions that are about to be processed through \qtwon\ queue} The miner can always process a transaction of lower value, say transaction $a$, than it should from the \qonen\ queue, by swapping out a transaction of higher value, say transaction $b$. Please refer to Fig. \ref{fig:swap}.  The miner might be inclined to do this if $a$ is about to be processed via the \qtwon\ queue, whereas $b$ is not. The miner might want to do this if he notices that $a$ is, in fact, of higher value than what is typically included in the \qonen\ section. In this sense, the miners keep $b$ held out, to be swapped in later (finally) for another transaction, say transaction $c$, that is lower in value as compared to $b$ and $a$, thus realizing a profit. The miner may swap several times (swapping out transactions $b$, $b'$, $b''$ ...) before finally swapping out transaction $c$. 

Note that during this attempt, it is easy to see that the miner is risking losing a profit as he is betting on finding the transaction $c$. He may not find such a transaction if the lowest value of the transactions included in the \qonen\ section never drops sufficiently, or the miner does not maintain a mining monopoly, and some other miner mines the next block gaining the risked amount.

The theoretical analysis of this attack strategy is not tractable, and nor is the simulating the attack feasible due to the complexity of the actions available and state-space. Thus, to show that this attack is ineffective, we give the adversary generous and impractical advantages and show that even in the best case of executing this attack in the backdrop of our simulation, the adversary gains as little as less than $1\%$ of the total rewards gained.

Since the transaction $a$ must ultimately be swapped with a transaction $c$, we simply consider the number of potentially ultimately successful swaps, as the minimum of; the number of transactions processed by the \qonen\ queue, that are valued below the maximum value of the transactions processed by the \qtwon\ queue; and the number of transactions processed by the \qtwon\ queue, that are valued above the minimum value of the transactions processed by the \qonen\ queue. We multiply the value obtained by the difference between; the maximum value of the transactions processed by the \qtwon\ queue and the minimum value of the transactions processed by the \qonen\ queue. In our simulations, the resultant value turns out to be less than $0.59 \pm 0.29\%$ of the total value of transactions processed.

Now, this is clearly an over-valuation for the following reasons;
(i) The risk of the attempt is completely ignored here.
(ii) The number of ultimately successful swaps considered is the result of the best (perhaps better than the practical best) possible exploitation of the one-to-one correspondence of the swapped transactions by the adversary.
(iii) The value of profit gained with each successful swap is just taken to be the maximum profit gained from the best possible swap.
(iv) The attacking miner is assumed to have a monopoly over mining, i.e., the attacking miner is the only one mining and hence can carry out this attack unhindered, i.e., without the possibility of another miner publishing a block impeding the attacking efforts.

\subsection{Discussion}
The parameters of \op, $\alpha$ and $f^0_{min}$ can be chosen by consensus and should be agreeable by both the miners and the users; we suggest $\alpha=0.2$ and $f^0_{min}=\frac{\mbox{average cost of mining a block}}{bs_{max}}$. If $\alpha=0$, then \op\ reduces to \btc, whereas $\alpha=1$ would be strictly \fifo. \btc, as we have seen, is not fair. Strictly \fifo\ processing disables the ability of the users to express urgency, and the average processing latency will not decrease with increasing $\eta$, which is a requirement for fairness for the users. An $f^0_{min}$ too low will discourage mining, an $f^0_{min}$ too high will discourage usage of the blockchain.

While we have chosen specific functions and parameters for $\phi$ and $\psi$, we believe that any monotonically increasing function for $\phi$ and any monotonically decreasing function for $\psi$ would yield similar yet scaled results.

\section{Conclusion}
In this paper, we studied \btc\ under \emph{mature operating conditions} (MOC), i.e., in TFOM and standard influx. To study a given blockchain, we introduced notions of fairness (i) for the miners and the users, and (ii) the health of a blockchain. Under reasonable assumptions, we showed using simulations that miners act greedily in \btc, as it is \ede\ to do so, and as a consequence, \btc\ ecosystem is not fair. To achieve fairness in \btc, we propose \op, a simple yet powerful modification to \btc. In \op; each transaction must include a minimum amount, to ensure that transaction fees cover marginal mining costs under MOC; and must have a minimum number of transactions offering the minimum specified amount. We showed using simulation analysis that in \op, miners act in favor of the health of the blockchain as it is \ene, and as a consequence, \op\ ecosystem is fair.

\bibliographystyle{IEEEtran}  % do not change this line!
\bibliography{ref}  % put name of your .bib file here

\begin{thebibliography}{10}
\providecommand{\url}[1]{#1}
\csname url@samestyle\endcsname
\providecommand{\newblock}{\relax}
\providecommand{\bibinfo}[2]{#2}
\providecommand{\BIBentrySTDinterwordspacing}{\spaceskip=0pt\relax}
\providecommand{\BIBentryALTinterwordstretchfactor}{4}
\providecommand{\BIBentryALTinterwordspacing}{\spaceskip=\fontdimen2\font plus
\BIBentryALTinterwordstretchfactor\fontdimen3\font minus
  \fontdimen4\font\relax}
\providecommand{\BIBforeignlanguage}[2]{{%
\expandafter\ifx\csname l@#1\endcsname\relax
\typeout{** WARNING: IEEEtran.bst: No hyphenation pattern has been}%
\typeout{** loaded for the language `#1'. Using the pattern for}%
\typeout{** the default language instead.}%
\else
\language=\csname l@#1\endcsname
\fi
#2}}
\providecommand{\BIBdecl}{\relax}
\BIBdecl

\bibitem{nakamoto2008bitcoin}
S.~Nakamoto \emph{et~al.}, ``Bitcoin: A peer-to-peer electronic cash system,''
  2008.

\bibitem{carlsten2016instability}
M.~Carlsten, H.~Kalodner, S.~M. Weinberg, and A.~Narayanan, ``On the
  instability of bitcoin without the block reward,'' in \emph{Proceedings of
  the 2016 ACM SIGSAC Conference on Computer and Communications
  Security}.\hskip 1em plus 0.5em minus 0.4em\relax ACM, 2016, pp. 154--167.

\bibitem{blockchain.com}
blockchain.com, ``Mempool transaction count,''
  \url{https://www.blockchain.com/charts/}, 2019, accessed: 15-November-2019.

\bibitem{bitcoinfees}
earn.com, ``Bitcoin fees for transactions,'' https://bitcoinfees.earn.com/,
  2019, accessed: 15-November-2019.

\bibitem{moser2015trends}
M.~M{\"o}ser and R.~B{\"o}hme, ``Trends, tips, tolls: A longitudinal study of
  bitcoin transaction fees,'' in \emph{International Conference on Financial
  Cryptography and Data Security}.\hskip 1em plus 0.5em minus 0.4em\relax
  Springer, 2015, pp. 19--33.

\bibitem{li2018transaction}
J.~Li, Y.~Yuan, S.~Wang, and F.-Y. Wang, ``Transaction queuing game in bitcoin
  blockchain,'' in \emph{2018 IEEE Intelligent Vehicles Symposium (IV)}.\hskip
  1em plus 0.5em minus 0.4em\relax IEEE, 2018, pp. 114--119.

\bibitem{houy2014economics}
N.~Houy, ``The economics of bitcoin transaction fees,'' \emph{GATE WP}, vol.
  1407, 2014.

\bibitem{easley2019mining}
D.~Easley, M.~O'Hara, and S.~Basu, ``From mining to markets: The evolution of
  bitcoin transaction fees,'' \emph{Journal of Financial Economics}, 2019.

\bibitem{kroll2013economics}
J.~A. Kroll, I.~C. Davey, and E.~W. Felten, ``The economics of bitcoin mining,
  or bitcoin in the presence of adversaries,'' in \emph{Proceedings of WEIS},
  vol. 2013, 2013, p.~11.

\bibitem{huberman2019economic}
G.~Huberman, J.~Leshno, and C.~C. Moallemi, ``An economic analysis of the
  bitcoin payment system,'' \emph{Columbia Business School Research Paper}, no.
  17-92, 2019.

\bibitem{kasahara2016effect}
S.~Kasahara and J.~Kawahara, ``Effect of bitcoin fee on
  transaction-confirmation process,'' \emph{arXiv preprint arXiv:1604.00103},
  2016.

\bibitem{koops2018predicting}
D.~Koops, ``Predicting the confirmation time of bitcoin transactions,''
  \emph{arXiv preprint arXiv:1809.10596}, 2018.

\bibitem{rizun2015transaction}
P.~R. Rizun, ``A transaction fee market exists without a block size limit,''
  \emph{Block Size Limit Debate Working Paper}, 2015.

\bibitem{bitcoin_magazine_2019}
B.~Magazine, ``What is the bitcoin block size limit?''
  \url{https://bitcoinmagazine.com/guides/what-is-the-bitcoin-block-size-limit},
  2019, accessed: 15-November-2019.

\bibitem{zou2010tolerable}
J.~Zou, S.~Gujar, and D.~Parkes, ``Tolerable manipulability in dynamic
  assignment without money,'' in \emph{Twenty-Fourth AAAI Conference on
  Artificial Intelligence}, 2010.

\end{thebibliography}

\end{document}